\documentclass[aps,prd,preprint,floatfix,nofootinbib,superscriptaddress]{revtex4}

\usepackage{graphicx,subfigure,epsfig,epsf}

\newcommand{\s}{\sigma}
\newcommand{\g}{\gamma}

\newcommand{\dU}{$d_{\mathcal{U}}$}
\newcommand{\LdaU}{$\Lambda_{\mathcal{U}}$}
\newcommand{\MU}{$M_{\mathcal{U}}$}
\newcommand{\OU}{$O_{\mathcal{U}}$}
\newcommand{\beq}{\begin{equation}}
\newcommand{\eeq}{\end{equation}}
\newcommand{\bea}{\begin{eqnarray}}
\newcommand{\eea}{\end{eqnarray}}

\def\Re{{\cal R \mskip-4mu \lower.1ex \hbox{\it e}\,}}
\def\Im{{\cal I \mskip-5mu \lower.1ex \hbox{\it m}\,}}

\def\etal{{\it et al.}}

\def\tev{\,{\ifmmode\mathrm {TeV}\else TeV\fi}}
\def\gev{\,{\ifmmode\mathrm {GeV}\else GeV\fi}}
\def\mev{\,{\ifmmode\mathrm {MeV}\else MeV\fi}}
\def\to{\rightarrow}

\begin{document}


\def\issue(#1,#2,#3){#1 (#3) #2} 
\def\APP(#1,#2,#3){Acta Phys.\ Polon.\ \issue(#1,#2,#3)}
\def\ARNPS(#1,#2,#3){Ann.\ Rev.\ Nucl.\ Part.\ Sci.\ \issue(#1,#2,#3)}
\def\CPC(#1,#2,#3){comp.\ Phys.\ comm.\ \issue(#1,#2,#3)}
\def\CIP(#1,#2,#3){comput.\ Phys.\ \issue(#1,#2,#3)}
\def\EPJC(#1,#2,#3){Eur.\ Phys.\ J.\ C\ \issue(#1,#2,#3)}
\def\EPJD(#1,#2,#3){Eur.\ Phys.\ J. Direct\ C\ \issue(#1,#2,#3)}
\def\IEEETNS(#1,#2,#3){IEEE Trans.\ Nucl.\ Sci.\ \issue(#1,#2,#3)}
\def\IJMP(#1,#2,#3){Int.\ J.\ Mod.\ Phys. \issue(#1,#2,#3)}
\def\JHEP(#1,#2,#3){J.\ High Energy Physics \issue(#1,#2,#3)}
\def\JPG(#1,#2,#3){J.\ Phys.\ G \issue(#1,#2,#3)}
\def\MPL(#1,#2,#3){Mod.\ Phys.\ Lett.\ \issue(#1,#2,#3)}
\def\NP(#1,#2,#3){Nucl.\ Phys.\ \issue(#1,#2,#3)}
\def\NIM(#1,#2,#3){Nucl.\ Instrum.\ Meth.\ \issue(#1,#2,#3)}
\def\PL(#1,#2,#3){Phys.\ Lett.\ \issue(#1,#2,#3)}
\def\PRD(#1,#2,#3){Phys.\ Rev.\ D \issue(#1,#2,#3)}
\def\PRL(#1,#2,#3){Phys.\ Rev.\ Lett.\ \issue(#1,#2,#3)}
\def\PTP(#1,#2,#3){Progs.\ Theo.\ Phys. \ \issue(#1,#2,#3)}
\def\RMP(#1,#2,#3){Rev.\ Mod.\ Phys.\ \issue(#1,#2,#3)}
\def\SJNP(#1,#2,#3){Sov.\ J. Nucl.\ Phys.\ \issue(#1,#2,#3)}



\bibliographystyle{revtex}

\title{Unparticle effects in Supernovae cooling} 




\author{Prasanta Kumar Das}
\affiliation{Birla Institute of Technology and Science-Pilani, Goa Campus, NH-17B, Zuarinagar, GOA- 403726}

\date{\today}

\begin{abstract}
Recently H.~Georgi suggested that a scale invariant unparticle ${\mathcal{U}}$ sector with an infrared fixed point at high energy can couple with the SM matter via a higher-dimensional operator suppressed by a high cut-off scale. Intense phenomenological search of this unparticle sector in the collider and flavour physics context has already been made. Here we explore it's impact in cosmology, particularly it's possible role in the supernovae cooling. We found that the energy-loss rate (and thus the cooling) is strongly dependent on the effective scale \LdaU  ~and the anomalous dimension \dU~ of this unparticle theory. 
\end{abstract}


\maketitle


\section{Introduction}
Scale invariance is an important concept in particle physics.
Although it is broken due to the masses of the particles of the standard model(SM), it gets restored at high energies much beyond the SM. Recently H.~Georgi\cite{Georgi1,Georgi2} suggested that a scale invariant non-trivial theory with an infra-red fix point can couple with the SM fields via an effective higher-dimensional operators suppressed by a high cut-off scale. Since the theory is scale invariant, it can not be described by the conventional particles(having mass), but by the conformal matter which Georgi termed as 'unparticles(${\mathcal{U}}$)' with the scale dimension \dU. Such an unparticle extension of the SM will be drastically  different from other type of SM extension e.g. supersymmetry, extra dimension from the phenomenological point of view and the differences can be set at the TeV energies. Several interesting unparticle phenomenology in collider, flavour physics and cosmology context are available in the literature \cite{Cheung,Luo,Chen,Ding,Liao,Aliev,Li,Duraisamy,Lu,Okada,Mohanta,
Davoudiasl,Huang,Freitas,ChenGeng,Anchordoqui,Majhi,McDonald,
SDas}.

The supernovae(SNe), a massive supergiant object of 10 billion solar mass, when explodes through the core collapse mechanism(e.g. SN1987A), releases huge energy  $3\times10^{53}$~ erg, out of which about 99\% going into neutrinos, whereas 1\% into the gravitational binding energy. Numerical neutrino light curves can be compared with the SN1987A data where the measured energies are found to be ``too low''. This raises the possibility whether SN1987A can loose it's energy by  other channels. Here we propose some novel channel which comprises the production of unparticle(in the form of missing energy) through which the core of the supernova can lose energy and thus cool. 
 
Before to explore this possibility, let us first review the basics of this unparticle physics as outlined in Ref.~\cite{Georgi1}.
  
\section{Unparticle physics}
The typical interaction of the SM fields with the Banks-Zaks(BZ)\cite{BZ} fields with an infra-red fixed point at high energy takes the form 
\bea
{\mathcal{L}}_{BZ} =\frac{O_{SM} O_{BZ}}{{M_{\mathcal{U}}}^k},~k > 0
\eea 
where \MU ~is the mass of the heavy exchange particle, $O_{SM}$ is the SM operator of mass dimension $d_{SM}$ and $O_{BZ}$ is the BZ operator of mass dimension $d_{BZ}$. Just like 
the non-abelian gauge theory, the renormalization effects in the scale invariant BZ sector causes dimensional transmutation in the BZ sector at an energy scale \LdaU ~and below this scale the scale-invariant BZ operator must match onto the unparticle operator of the form 
\bea \label{eqn:SMU}
\frac{C_{\mathcal{U}}{{\Lambda_{{\mathcal{U}}}}}^{d_{BZ} -d_{\mathcal{U}} }}{{M_{\mathcal{U}}}^k} O_{SM} O_{{\mathcal{U}}}  
\eea
where \dU ~is the scale dimension of the unparticle operator \OU ~and~ $C_{\mathcal{U}}$ is the coefficient function fixed by matching and is different for different unparticle, not universal.
 In Ref.\cite{Georgi1}, Georgi pointed out three possible structures of unparticle operators: $O_{\mathcal{U}}$,~${O_{\mathcal{U}}}^\mu$ and ${O_{\mathcal{U}}}^{\mu \nu}$.
They are taken to be hermitian and operators ${O_{\mathcal{U}}}^\mu$ and ${O_{\mathcal{U}}}^{\mu \nu}$ are assumed to be transverse. The typical structure of the effective operators(of the form of Eq.~\ref{eqn:SMU}) which are phenomenologically interesting are 
\bea
\lambda_0 \frac{1}{\Lambda_{{\mathcal{U}}}^{d_{{\mathcal{U}}}}}G_{\rho \sigma} G^{\rho \sigma} O_{\mathcal{U}}, ~~ \lambda_1 \frac{1}{\Lambda_{{\mathcal{U}}}^{d_{{\mathcal{U}}}-1}} {\overline{f}} \gamma_\mu f O_{\mathcal{U}}^\mu,~~and~
\lambda_2 \frac{1}{\Lambda_{{\mathcal{U}}}^{d_{{\mathcal{U}}}}}G_{\rho \alpha} 
G_{\sigma}^{\alpha} O_{\mathcal{U}}^{\rho \sigma},
\eea  
where $G_{\rho \sigma}$ stands for the photon, gluon field strength, 
$f$, the SM fermion and $\lambda_i(i=0,1,2)$ are the dimensionless effective coupling constants $\frac{C_{\mathcal{U}}{{\Lambda_{{\mathcal{U}}}}}^{d_{BZ}}}{{M_{\mathcal{U}}}^k}$. In the present anaylsis, however, we will confine ourselves in the scalar operator coupling with the photon field i.e. of the first kind of the above set.

 In Ref.~\cite{Georgi1} Georgi also showed that $d\Phi$, the phase space of an unparticle operator of dimension \dU ~is the same as the phase space 
of n=\dU ~massless invisible particles and also  
\dU ~need not necessarily be an integer. 
Georgi showed that \cite{Georgi1} the unparticle phase space $d\Phi$ is    
 proportional to the coefficient function
\bea \label{Adu}
A_{d_{{\mathcal{U}}}}=\frac{16 \pi^{5/2}}{(2 \pi)^{2 d_{{\mathcal{U}}}}} 
\frac{\Gamma(d_{{\mathcal{U}}} + 1/2)}{\Gamma(d_{{\mathcal{U}}} - 1) \Gamma(2 d_{{\mathcal{U}}})}.
\eea
Following the reason  exemplified in Ref.~\cite{Georgi2}, we will assume $ 1 \le d_{{\mathcal{U}}} \le 2$ in our analysis.

\section{Supernova Explosion and Cooling}
Supernovae come in two main observational varieties:
Type II are those whose optical spectra exihibit hydrogen lines and have less sharp peaks at maxima (of 1 billion solar luminosities), whereas the optical spectra for the Type I supernovae does not have any hydrogen lines and it exhibits sharp maxima \cite{VHS}.
Physically, there are two fundamental types of supernovae, based on what mechanism powers them: the thermonuclear SNe and the core-collapse ones. Only SNe Ia are thermonuclear type and the rest are formed by core-collapse of a massive star. 
The core-collapse supernovae are the class of explosions which mark the evolutionary end of massive stars ($M \geq 8\,M_\odot$). 
 The kinetic energy of the explosion carries about 1\% of the liberated gravitational binding energy of about 
$3\times10^{53}~{\rm ergs}$ and the remaining 99\% going into neutrinos. This powerful and detectable neutrino burst is the main astro-particle interest of core-collapse SNe.

In the case of SN1987A, about $10^{53}$ ergs of gravitational binding energy was released in few seconds and the neutrino fluxes were measured by Kamiokande \cite{Kamio} and IMB \cite{IMB} collaborations. Numerical neutrino light curves can be compared with the SN1987A data where the measured energies are found to be ``too low''.  For example,
the numerical simulation in \cite{Totani:1997vj} yields time-integrated values $\langle E_{\nu_e}\rangle\approx13~{\rm MeV}$, $\langle E_{\bar\nu_e}\rangle\approx16~{\rm MeV}$, and $\langle E_{\nu_x}\rangle\approx23~{\rm MeV}$.  On the other hand, the data imply $\langle E_{\bar\nu_e}\rangle=7.5~{\rm MeV}$ at Kamiokande and 11.1~MeV at IMB~\cite{Jegerlehner:1996kx}.  Even the 95\% confidence range for Kamiokande implies $\langle E_{\bar\nu_e}\rangle<12~{\rm MeV}$.  Flavor oscillations would increase the expected energies and
thus enhance the discrepancy~\cite{Jegerlehner:1996kx}.  It has remained unclear if these and other anomalies of the SN1987A neutrino signal should be blamed on small-number statistics, or point to a serious problem with the SN models or the detectors, or is there a new physics happening in SNe?

Since we have these measurements already at our disposal, now if we propose some novel channel through which the SNe core of the supernova can lose energy, the luminosity in this channel should be low enough to preserve the agreement of neutrino observations with theory. That is,
${\cal L}_{new\, channel} \leq 10^{53}\, ergs\, s^{-1}.$
This idea was earlier used to put the strongest experimental upper bounds on the axion mass \cite{axions}. Here, we will consider the emission of unparticles 
from the SNe core which will carry the energy(say missing energy) with them.  The constraint on luminosity of this process can be converted into a bound on the \dU ~and~ \LdaU. Any mechanism which leads to significant energy-loss from the SNe core immediately after bounce will produce a very different neutrino-pulse shape, and so will destroy this agreement, which in the case of axion is explicitly shown by Burrows's \etal~\cite{BBT}.
Raffelt has proposed a simple analytic criterion based on detailed supernova simulations~\cite{Raffelt}: if any energy-loss mechanism has an emissivity greater than $10^{19}$ ergs g$^{-1}$ s$^{-1}$ then it will remove sufficient energy from the explosion to invalidate the current understanding of Type-II supernovae's neutrino signal. 

\section{Unparticle production and supernovae cooling}
The unparticle role in the SN 1987A cooling have been looked in detail and are available in the literature \cite{Davoudiasl} and \cite{Hannestad}). 
In all such studies, the unparticle interaction with nucleons, neutrinos, electrons is considered, but with the photon is not taken into account. As we will see here that the photon (which is quite abundant inside SNe although it's density is somewhat less than the matter density) interaction with the unparticle stuff may be quite important in this SNe cooling and thus put stringent bound on unparticle parameters \dU ~and \LdaU.  
The task is to find first the unparticle production cross-section in 
photon-photon fushion inside SN 1987A. Before to do that let us recall the two-point correlation function of the scalar like 
unparticle operator $O_{\mathcal{U}}$ \cite{Georgi1}
\bea
\langle O|O_{\mathcal{U}}(x) O^\dagger_{\mathcal{U}}(0)|O \rangle 
= \int \frac{d^4 P}{(2 \pi)^4} e^{-i P.x} \left|\langle O|O_{\mathcal{U}}(O) |P \rangle\right|^2 \rho(P^2)
\eea

\noindent with $\left|\langle O|O_{\mathcal{U}}(O) |P \rangle\right|^2 \rho(P^2) = A_{d_{{\mathcal{U}}}} \theta(P^0) \theta(P^2) (P^2)^{d_{\mathcal{U}} - 2}$. The state
$|P\rangle$ corresponds to the unparticle state with momentum $P^\mu$. 

Now photons are quite abundant in supernovae. So phenomenologically an interesting process responsible for supernovae cooling might be the photon-photon annihilation to unparticles i.e.
\begin{equation}
\gamma(k_1) + \gamma(k_2) \to {\mathcal{U}}(P).
\end{equation} 
It is now straightforward to find the cross section of the above process and found to be  
\bea \label{CSggU}
\s_{\g \g \to {\mathcal{U}}}(S,d_{\mathcal{U}},\Lambda_{\mathcal{U}}) = \frac{1}{8} \left( \frac{\lambda_o}
{\Lambda_{\mathcal{U}}^{d_{\mathcal{U}}}}\right)^2 A_{d_{{\mathcal{U}}}} ~\theta(P^0) ~\theta(P^2)~  S^{d_{\mathcal{U}} - 1},
\eea
where $S = (k_1 + k_2)^2 = P^2$. Here we assume that two interacting photon are purely transverse in nature. The  plasmon effect(as a result of which the photon can achive the longitudinal degree of freedom and become massive) inside supernovae \cite{Raffelt} is not considered here.
Since we are concerned with the energy loss to unparticles as missing energies, it is convenient and standard
\cite{Kolb} to define the quantity $\dot{\epsilon}_{\g + \g \to {\mathcal{U}}}$
which stands for the rate at which energy is lost to unparticles 
via the process $\g + \g \to {\mathcal{U}} $, per unit time per unit mass of the stellar object. In terms of the cross-section $\sigma_{\g + \g \to {\mathcal{U}}}$,
the photon number densities  $n^{(1)}_\g$ and $n^{(2)}_\g$ and the mass density
$\rho$, $\dot{\epsilon}$ is given by
\begin{equation}
\dot{\epsilon}_{\g + \g \to {\mathcal{U}}} = \frac{\langle n^{(1)}_\g n^{(2)}_\g
\sigma_{(\g + \g \to {\mathcal{U}})} v_{rel} E_{cm} \rangle}{\rho}
\label{emrate}
\end{equation} 
where the brackets indicate thermal averaging, $E_{cm}(=E_1 + E_2)$, the center of mass energy and $v_{rel}=\mathbf{p} \sqrt{S}/(E_a E_b)$, where $\mathbf{p}=\mathbf{p}_1 =\mathbf{p}_2 = \frac{\lambda^{1/2}(s,m_1^2,m_2^2)}{2 \sqrt{s}}$ in the c.o.m frame of two photons. The function $\lambda(x,y,z)(=x^2 + y^2 + z^2 - 2 x y -2 y z - 2 z x)$, is the standard  $K{\ddot{a}}$llen function.

Assuming photons inside supernovae of temperature $T$ follows Bose-Einstein distribution, the above energy loss rate can be explicitly written as 
\bea \label{lossrate1}
\dot{\epsilon}_{\g + \g \to {\mathcal{U}}} = \frac{1}{\rho} \int {2 d^3{\vec k}_1\over (2\pi)^3} {1\over e^{E_1/T}-1}
\int {2 d^3{\vec k}_2\over (2\pi)^3} {1\over e^{E_2/T}-1}
{S (E_1+E_2)\over 2E_1E_2} \sigma_{\gamma \gamma \to {\mathcal{U}}}(S,d_{\mathcal{U}},\Lambda_{\mathcal{U}}).
\eea
where $\sigma_{\gamma \gamma \to {\mathcal{U}}}$ is given in Eq.~(\ref{CSggU}). Introducing two dimensionless variables $x_1= E_1/T$ and $x_2=E_2/T$, 
Eq.~(\ref{lossrate1}) can be re-written as
\bea \label{lossrate2}
\dot{\epsilon}_{\g + \g \to {\mathcal{U}}} =\frac{1}{16 \pi^4 \rho} \left(\frac{\lambda_0}{\Lambda_{\mathcal{U}}^{d_{\mathcal{U}}}}\right)^2~ A_{d_{\mathcal{U}}}
T^{2 d_{\mathcal{U}} + 5} 
\int_o^\infty d x_1 \frac{x_1}{e^{x_1}-1} ~\int_o^\infty d x_2 \frac{x_2}{e^{x_2}-1}
 (x_1 + x_2)^{2 d_{\mathcal{U}} + 1}.
\eea
where $A_{d_{\mathcal{U}}}$ is given in Eq.~(\ref{Adu}).

\section{Results and analysis}
We now analyze the supernovae energy loss rate to unparticles under certain criteria. The criteria is that the loss rate is less than equal to $10^{19}~ erg~ g^{-1} s^{-1}$. Using this we can impose constraints on the scale \LdaU~ and anomalous dimension \dU. In Figure 1a we have shown the energy loss rate to unparticles as a function of  the scale \LdaU~ for different \dU~ varying from $1.1$ to $1.9$. The following inputs: the supernovae temparature  $T=30$ MeV, $\lambda_0=10^{-3}$ and the supernovae core density $\rho \simeq 10^{15}$ $g~cm^{-3}$ 
\cite{Raffelt},we have taken into consideration while obtaining such plots. The horizontal line corresponding to the upper bound on the supernovae energy loss rate, gives rise the lower bound on \LdaU~ for different \dU. Note that with the increase of \dU~ the lower bound on \LdaU~ decreases. A plot showing such a  variation of \LdaU~ with \dU~ is shown in Figure 1b.  Note that in Figure 1b we have set $1 <$ \dU $< 2$ as was suggested by Georgi \cite{Georgi1, Georgi2}. The lower and the upper horizontal curves respectively, stands for \LdaU $= 1$ TeV and $10$ TeV. The region above the curve(undotted red one) of Figure 1b is allowed. Interestingly whatsoever be the form of the New Physics, required to get rid of the shortcomings/limitations of the standard model, it is widely believed to be around the TeV scale. If we take this point of view, then Figure 1b quite interestingly allows us to make one important prediction. One finds that for the effective scale \LdaU~(which is also the scale
\newpage
\vspace*{-0.5in}
\begin{figure}
\subfigure[]{
\label{PictureThreeLabel}
\hspace*{-0.7 in}
\begin{minipage}[b]{0.5\textwidth}
\centering
\includegraphics[width=\textwidth]{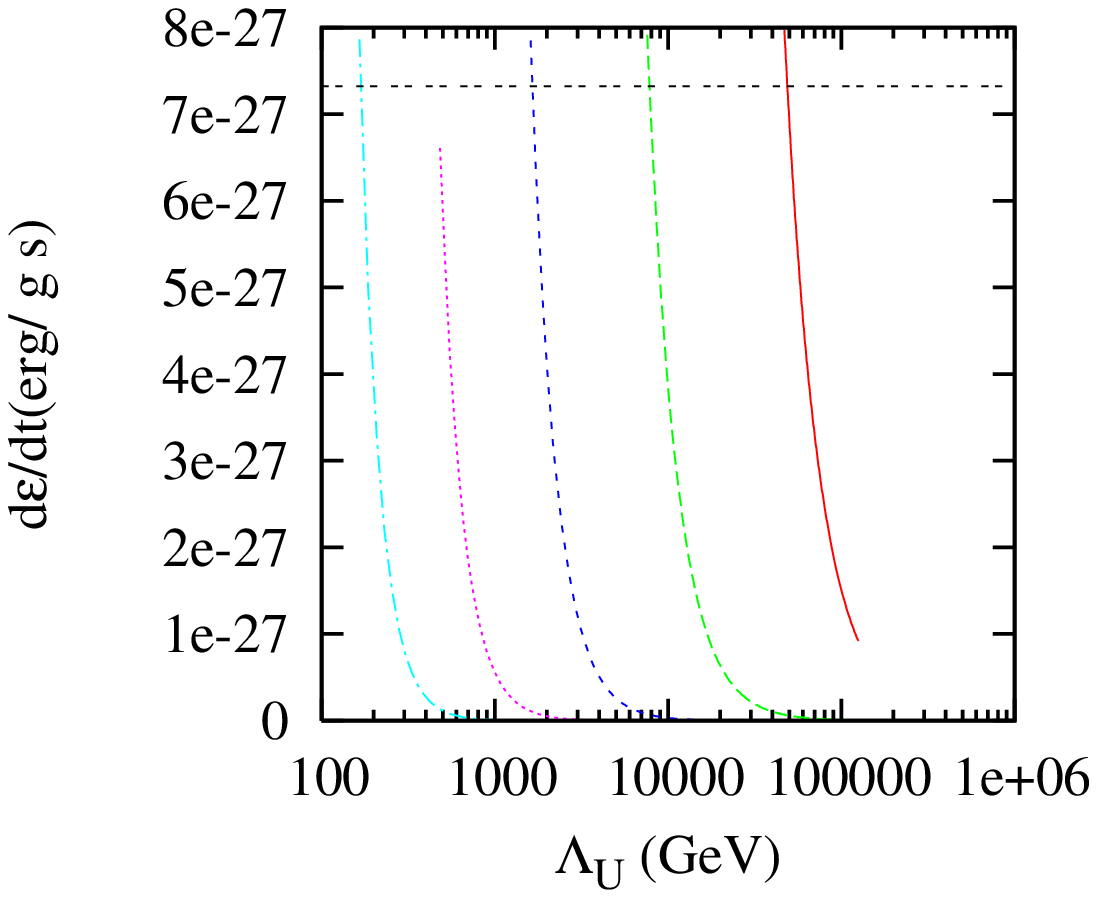}
\end{minipage}}
\subfigure[]{
\label{PictureFourLabel}
\hspace*{0.3in}
\begin{minipage}[b]{0.5\textwidth}
\centering
\includegraphics[width=\textwidth]{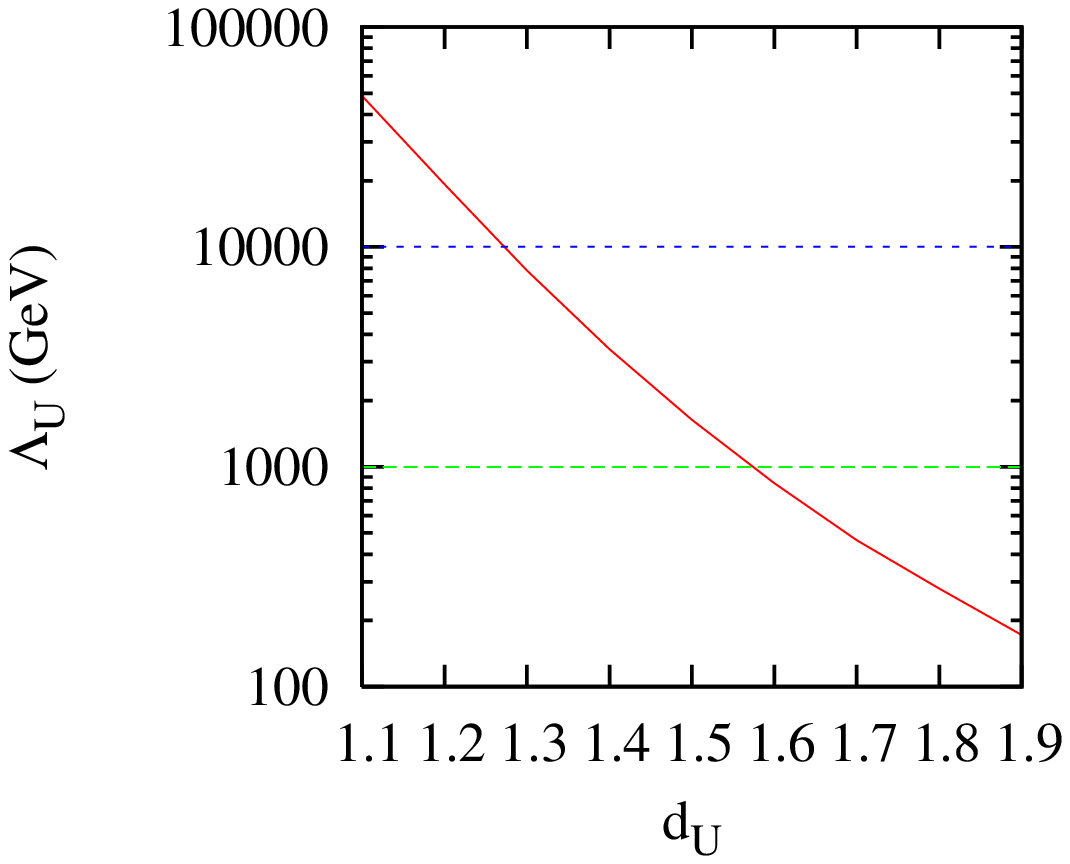}
\end{minipage}}
\end{figure}
\noindent {\bf Figure 1[a,b]}.
{{In Figure 1a, the supernovae energy loss rate $d\epsilon/dt$ ($erg~ g^{-1} s^{-1}$) as a function of \LdaU~ is shown. For the extreme right curve \dU=1.1, whereas going to left from right \dU~ increases by $0.2$. The extreme left curve  corresponds to \dU~=1.9. In Figure 1b, the lower bound on \LdaU~ as a function of \dU~ is shown which follows from the constraint $d\epsilon/dt < 10^{19}$ $erg~g^{-1}~s^{-1}$. The two horizontal curves in Figure 1b corresponds to \LdaU~=1 TeV and 10 TeV.}}

\noindent of New Physics) lying between $1$ to $10$ TeV, the anomalous dimension \dU~ of this scalar like unparticle stuff lies in between $\sim 1.3$ to $\sim 1.6$.
So if in near future New Physics appears in the form of missing energy, unpartcle(scalar-like) stuff with the above set of properties can be a strong candidate for that.

\section{{Discussions and Conclusions}}
We have considered the process photon photon annihilation to unparticles  as a possible mechanism for the energy-loss of the supernovae(SN1987). The lower bound on the effective scale \LdaU~of the unparticle stuff is obtained by requiring that the energy-loss rate $\le 10^{19}~ergs~g^{-1}~s^{-1}$ and we saw that it is quite sensitive of the anomalous dimension  \dU~ of the unparticle stuff: with the increase of \dU, the lower bound on \LdaU~ decreases. Interestingly we find that for the effective scale $1$ TeV$ <$ \LdaU~ $< 10$ TeV, the anomalous dimension \dU~ lies in between $\sim 1.3$ and $\sim 1.6$.

\section{{Acknowledgement}}
The author would like to thank Prof. Ramesh Kaul of IMSc,chennai for his useful comments. He would also like to acknowledge the collaboration with Satheesh Kumar and Dr. P. K. Suresh in a similar project but in context of large extra dimension.

\end{document}